\documentclass[onecollarge,natbib]{svjour2}
\bibpunct{[}{]}{;}{n}{}{,} 
\smartqed  
\usepackage{graphicx}
\usepackage{hepunits}
\usepackage{tikz}
\usetikzlibrary{arrows,snakes,backgrounds}
\usetikzlibrary{arrows}

\newcommand{\ImH}{Im \mathcal{H}}
\newcommand{\ReH}{Re \mathcal{H}}
\journalname{Few-Body Systems}
\begin{document}

\title{On Deeply Virtual Compton Scattering at next-to-leading order\thanks{Work partly supported by the Polish Grants NCN No DEC-2011/01/D/ST2/02069, by the Joint Research Activity "Study of Strongly Interacting Matter" (acronym HadronPhysics3, Grant Agreement n.283286) under the Seventh Framework Programme of the European Community, by the GDR 3034 PH-QCD, and the ANR-12-MONU-0008-01, and by the COPIN-IN2P3 Agreement.}
}


\author{H.~Moutarde         \and
        B.~Pire        \and
        F.~Sabati\'e        \and
        L.~Szymanowski        \and
        J.~Wagner 
}


\institute{H.~Moutarde, Irfu/SPhN, CEA, Centre de Saclay, F91191 Gif-sur-Yvette, France. \\ \email{herve.moutarde@cea.fr}. \\ 
           \and
           B.~Pire, CPHT, {\'E}cole Polytechnique, CNRS, 91128 Palaiseau, France. \\
	\and
	F.~Sabati\'e, Irfu/SPhN, CEA, Centre de Saclay, F91191 Gif-sur-Yvette, France. \\
	\and
	L.~Szymanowski, National Centre for Nuclear Research (NCBJ), Warsaw, Poland. \\
	\and
	J.~Wagner, National Centre for Nuclear Research (NCBJ), Warsaw, Poland.
}

\date{Received: date / Accepted: date}

\maketitle

\begin{abstract}
Deeply Virtual Compton Scattering in the near forward kinematic region is the golden access to Generalized Parton Distributions. We studied the $\mathcal{O}(\alpha_S)$ corrections to the scattering amplitude for both spacelike and timelike kinematics relevant respectively to the leptoproduction of a real photon and to the photoproduction of a lepton pair. It turns out that these corrections are phenomenologically important and that the gluonic contributions are by no means negligible, even in the moderate energy range of JLab12 and of the COMPASS-II experiment at CERN.
\end{abstract}


\section{Introduction}

Generalized Parton Distributions (GPDs) have been introduced independently by Mueller \textit{et al.} \cite{Mueller:1998fv}, Radyushkin \cite{Radyushkin:1997ki} and Ji \cite{Ji:1996ek}. They provide unique information about the 3D structure and the spin structure of the nucleon. They have been continuously at the heart of an intense theoretical and experimental activity as can be testified by the different reviews of this field \cite{Goeke:2001tz, Diehl:2003ny, Belitsky:2005qn, Boffi:2007yc, Guidal:2013rya}. 

Deeply Virtual Compton Scattering (DVCS) immediately appeared as the most promising channel to access GPDs \cite{Ji:1996ek}. Its crossed process, Timelike Compton Scattering (TCS), attracted attention later \cite{Berger:2001xd} and, even if the experimental situation is not as mature as in the DVCS case, much progress \cite{NadelTuronski:2009zz, Horn:2011zz} is expected in forthcoming years.

With the Jefferson Lab upgrade at 12~$\GeV$ and the beginning of the COMPASS-II experiment, the field of GPDs will enter an era of unprecedented precision. In this work we explore the consequences of the inclusion of Next-to-Leading Order (NLO) gluon coefficient functions and NLO corrections to the quark coefficient functions entering both DVCS and TCS amplitudes. Firstly, we compute spacelike and timelike Compton Form Factors (CFFs) with two models of GPDs. Secondly, we evaluate specific observables in kinematic conditions soon accessible in lepton nucleon collisions.


\section{Theoretical framework}


\subsection{Compton scattering}

\begin{figure*}
	\centering
	\includegraphics[width=\textwidth]{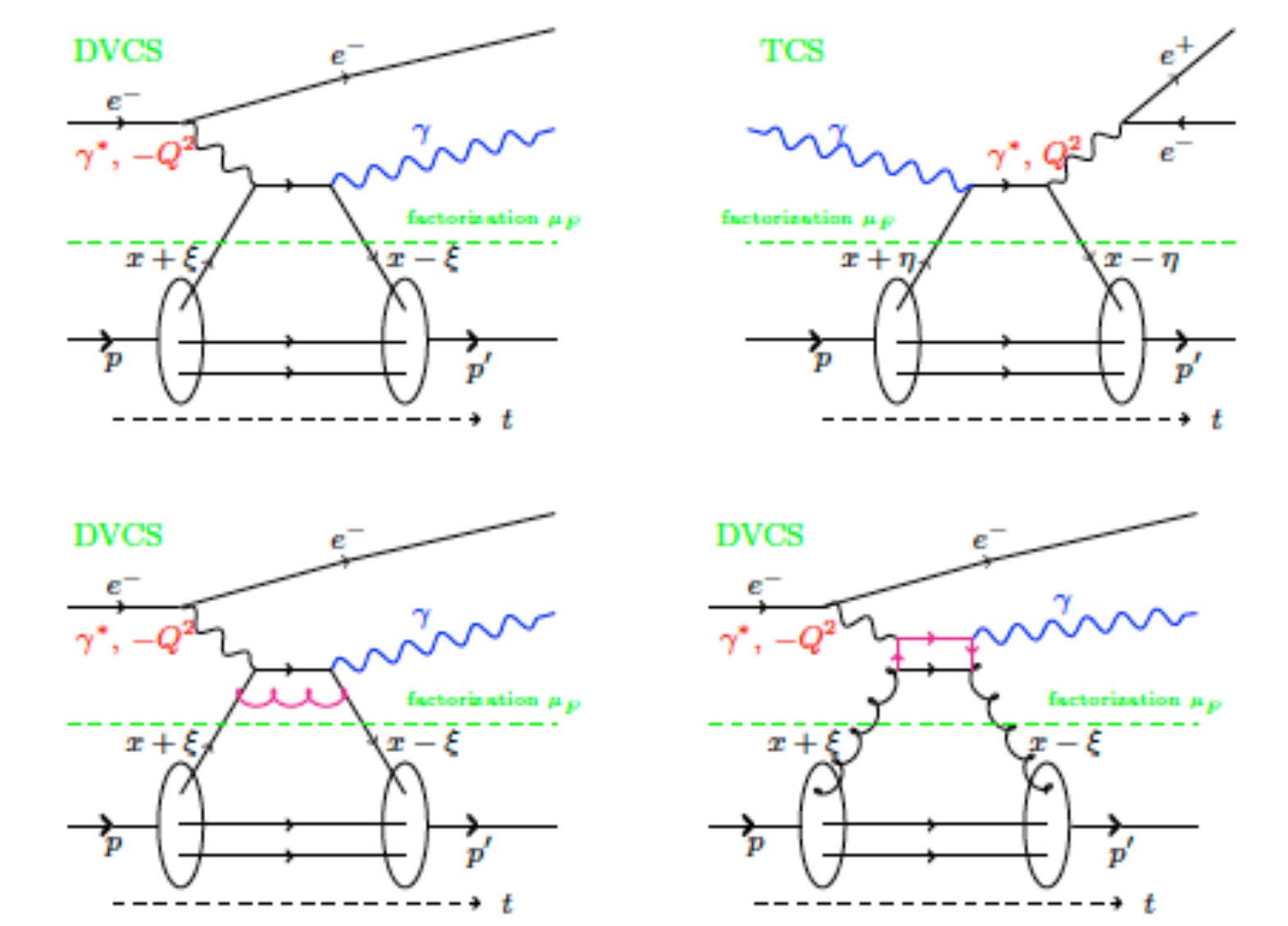}
	\caption{DVCS (upper line, left column) and TCS at LO (upper line, right column) and two contributions to DVCS at NLO involving quark (lower line, left column) and gluon (lower line, right column) GPDs.}
	\label{fig:Feynamn-diagrams-Compton-scattering-LO-and-NLO}       
\end{figure*}

Leading Order (LO) and some NLO contributions to the DVCS and TCS amplitudes are shown on Fig.~\ref{fig:Feynamn-diagrams-Compton-scattering-LO-and-NLO}. We denote the virtuality of the absorbed (DVCS) or emitted (TCS) virtual photon by $Q^2$ , the skewness variables by $\xi$ (DVCS) and $\eta$ (TCS),  the momentum transfer on the nucleon by $t = ( p - p' )^2$ and the factorization scale by $\mu_F$. See Ref.~\cite{Moutarde:2013qs} for detailed notations. At LO, only quark GPDs contribute to these processes, but NLO contributions include contributions due to quark and gluon GPDs as well.

\subsection{Explicit expressions}

NLO coefficient functions have been computed for the first time in Ref.~\cite{Belitsky:1997rh} for the DVCS case, and in Ref.~\cite{Pire:2011st} for the TCS case. A first-principle relation between the DVCS and TCS coefficient functions at NLO has been established in Ref.~\cite{Muller:2012yq}.

The Compton scattering amplitudes can be expressed in terms of CFFs. Generally speaking the quark CFF $\mathcal{H}_q$ is related to the unpolarized quark and gluon GPDs $H_q$ and $H_g$ by the following relation:
\begin{equation}
\mathcal{H}_q( \xi, Q^2 )
= \int_{-1}^{+1} dx \, H^+_q( x, \xi, \mu_F ) T_q\left(  x, \xi, \alpha_S( \mu_F ), \frac{Q}{\mu_F}  \right) +  \int_{-1}^{+1} dx \, H_g( x, \xi, \mu_F ) T_g\left(  x, \xi, \alpha_S( \mu_F ), \frac{Q}{\mu_F}  \right), 
\label{eq:def-CFF-H-all-orders}
\end{equation}
where we consider the singlet combination $H^+_q( x  ) = H_q( x  ) - H_q( - x  )$. Eq.~(\ref{eq:def-CFF-H-all-orders})  has the following simple form at LO expressed in terms of the Born coefficient function $C_0^q$:
\begin{equation}
\mathcal{H}_q( \xi, Q^2 ) 
 \stackrel{\textrm{LO}}{=} \int_{-1}^{+1} dx \, H^+_q( x, \xi, \mu_F ) C_0^q( x, \xi ), 
\label{eq:def-CFF-H-LO}
\end{equation}
and the more complex form at NLO:
\begin{eqnarray}
\mathcal{H}_q( \xi, Q^2 ) 
& \stackrel{\textrm{NLO}}{=} & \int_{-1}^{+1} dx \, H^+_q( x, \xi, \mu_F ) \left[ C_0^q( x, \xi ) + C_1^q\big(  x, \xi, \alpha_S( \mu_F )  \big) + \frac{1}{2} \ln \frac{| Q^2 |}{\mu_F^2} C_{\textrm{Coll}}^q\big(  x, \xi, \alpha_S( \mu_F ) \big)  \right] \nonumber \\
& & \quad +  \int_{-1}^{+1} dx \, H_g( x,\xi, \mu_F ) \left[ C_1^g\big(  x, \xi, \alpha_S( \mu_F ) \big) + \frac{1}{2} \ln \frac{| Q^2 |}{\mu_F^2} C_{\textrm{Coll}}^g\big(  x, \xi, \alpha_S( \mu_F ) \big) \right]  
\label{eq:def-CFF-H-NLO}
\end{eqnarray}
See Ref.~\cite{Moutarde:2013qs} for explicit expressions of the coefficient functions $C_0$, $C_1$ and $C_{\textrm{Coll}}$.

The differences between Eqs.~(\ref{eq:def-CFF-H-LO}) and (\ref{eq:def-CFF-H-NLO}) have some consequences regarding GPD extractions. Indeed the expression of the imaginary part of the CFF $\mathcal{H}_q$ changes from $\pi H^+_q( \xi, \xi, \mu_F )$ to:
\begin{eqnarray}
\ImH_q( \xi, Q^2 ) 
& \stackrel{\textrm{NLO}}{=} & \mathcal{I}( \xi ) H^+_q( \xi, \xi, \mu_F ) + \int_{-1}^{+1} dx \, T_q\left(  x, \xi, \alpha_S( \mu_F ), \frac{Q}{\mu_F}  \right)  \Big( H^+_q( x, \xi, \mu_F ) - H^+_q( \xi, \xi, \mu_F ) \Big) \nonumber \\
& & \quad + \textrm{ gluon contributions,}
\end{eqnarray}
where $\mathcal{I}( \xi )$ is a function of $\xi$. The integral probes the GPD $H_q$ at values of $x \neq \xi$ and the whole expression involve gluon contributions. There is no more direct link between the imaginary part of the CFF $\mathcal{H}_q$ and the value of the GPD $H_q$ on the cross-over line, even in the valence region where $H_q( -\xi, \xi)$ is expected to be small. While it is still possible and valuable to extract CFFs in an almost model-independent way along the lines of Ref.~\cite{Guidal:2008ie, Guidal:2009aa, Guidal:2010de, Guidal:2010ig, Kumericki:2013br}, the interpretation of the extracted CFF does not seem transparent anymore.


\section{Evaluation of Compton Form Factors}

In this section we evaluate DVCS and TCS CFFs with two GPD models based on Double Distributions (DDs) \cite{Mueller:1998fv, Radyushkin:1997ki, Musatov:1999xp}. DDs naturally achieve one of the strongest constraints on GPDs: the polynomiality of the Mellin moments of GPDs. They also automatically restore usual PDFs in the forward limit.


\subsection{Models of Generalized Parton Distributions}

\subsubsection{The Goloskokov~-~Kroll model}

The Goloskokov~-~Kroll (GK) model was developed in a series of papers \cite{Goloskokov:2005sd, Goloskokov:2006hr, Goloskokov:2007nt} to study Deeply Virtual Meson Electroproduction (DVMP). It has been recently tested againts DVCS data in a systematic way \cite{Kroll:2012sm}.

The GK model contains the following ingredients:
\begin{itemize}
	\item The Radyushkin DD factorized Ansatz (RDDA). For $i$ = $g$, sea or val:
\begin{eqnarray}
H_i( x, \xi, t ) & = & \int_{| \alpha | + | \beta | \leq 1} d\beta d\alpha \, \delta( \beta + \xi \alpha - x ) f_i( \beta, \alpha, t ), \\
f_i( \beta, \alpha, t ) & = & e^{b_i t} \frac{1}{| \beta |^{\alpha' t}} h_i( \beta ) \pi_{n_i}( \beta, \alpha ), \\
\pi_{n_i}( \beta, \alpha ) & = & \frac{\Gamma( 2 n_i + 2 )}{2^{2 n_i + 1} \Gamma^2( n_i + 1 )} \frac{( 1 - | \beta | )^2 - \alpha^2 ]^{n_i}}{( 1 - | \beta | )^{2 n_i + 1}}.
\end{eqnarray}

	\item The expressions for $h_i$ and $n_i$ are the following:
\begin{equation}
\begin{array}{lclclcl}
h_g( \beta ) & = & | \beta | g( | \beta | ) & \hspace{1cm} & n_g & = & 2, \\
h^q_{\textrm{sea}}( \beta ) & = & q_{\textrm{sea}}( | \beta | ) \textrm{sign}( \beta ) & \hspace{1cm} & n_{\textrm{sea}} & = & 2, \\
h^q_{\textrm{val}}( \beta ) & = & q_{\textrm{val}}( \beta ) \Theta( \beta ) & \hspace{1cm} & n_{\textrm{val}} & = &  1.
\end{array}
\end{equation}
\end{itemize}

This model is built from the CTEQ6m Parton Distribution Function (PDF) set \cite{Pumplin:2002vw}.

\subsubsection{The MSTW08-based GPD model}

To avoid drawing conclusions relying on a single GPD model, we built another GPD model in the RDDA framework:
\begin{itemize}
	\item We use MSTW08 Parton Distribution Functions \cite{Martin:2009iq}.

	\item We assume factorized $t$-dependence:
			\begin{equation}
H( x, \xi, t ) = \int_{| \alpha | + | \beta | \leq 1} d\beta d\alpha \, \delta( \beta + \xi \alpha - x ) \pi( \beta, \alpha ) f( \beta, t ).
			\end{equation}
			For $u$ and $d$ quarks:
\begin{eqnarray}
f_u( \beta, \alpha, t ) & = & \frac{1}{2} F^u_1( t ) u( \beta ) \pi( \beta, \alpha ). \\
f_d( \beta, \alpha, t ) & = & F^d_1( t ) d( \beta ) \pi( \beta, \alpha ).  
\end{eqnarray}
with $F_1^u$ and $F_1^d$ the $u$ and $d$ quark contributions to the proton form factor $F_1$. For $s$ quark and gluons a dipole Ansatz was used.

	\item We add a D-term coming from Chiral Quark Soliton Model (see Refs~\cite{Goeke:2001tz, Moutarde:2013qs} for details).
\end{itemize}


\subsection{Reminder: the GK model confronted to DVCS measurements}

Before using the GK model in order to compute CFFs on a wide kinematic range at LO and NLO, let us remind briefly the phenomenological successes and limitations of the model.

\begin{figure}
	\centering
	\includegraphics[width=0.45\textwidth]{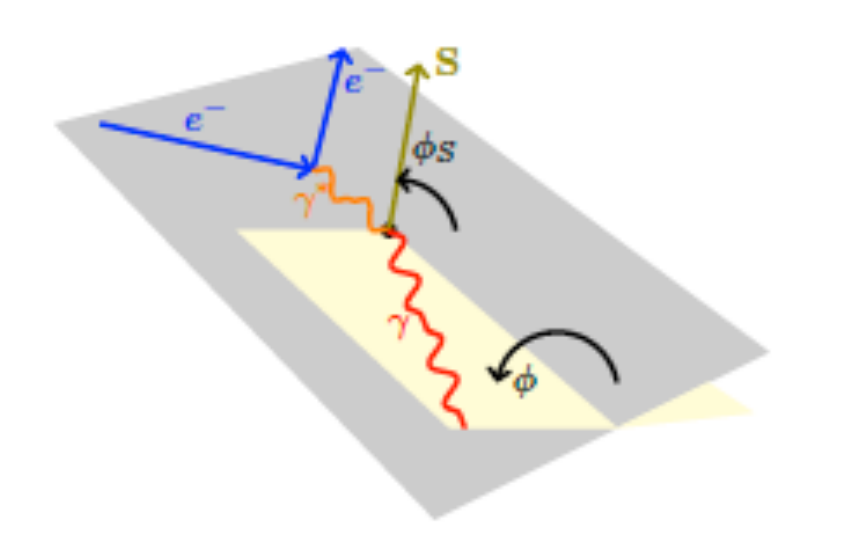}
	\caption{Kinematics of DVCS in the target rest frame.}
	\label{fig:DVCS-kinematics}
\end{figure}

Fig.~\ref{fig:DVCS-kinematics} pictures DVCS kinematics. The angle $\phi$ between the leptonic and hadronic planes obey the Trento convention \cite{Bacchetta:2004jz}. Following Ref.~\cite{Kroll:2012sm} we consider a lepton beam with helicity $h_e$ and charge $Q_e$ (in units of $|e|$) and define combined beam-spin and charge asymmetries:
\begin{equation}
d\sigma^{h_e,Q_e}(\phi) = d\sigma_{\rm UU}(\phi)\left[1 + h_e A_{\rm LU,DVCS}(\phi) + Q_e h_e A_{\rm LU,I}(\phi) + Q_e A_{\rm C}(\phi)\right].	
\end{equation}
Single beam-spin asymmetry can be defined as well:
\begin{equation}
A_{\rm LU}^{Q_e}(\phi)=\frac{d\sigma^{\stackrel{Q_e}{\rightarrow}} - d\sigma^{\stackrel{Q_e}{\leftarrow}}} {d\sigma^{\stackrel{Q_e}{\rightarrow}} + d\sigma^{\stackrel{Q_e}{\leftarrow}}}.	
\end{equation}
HERMES DVCS data are usually described through the Fourier coefficients of the various measured asymmetries.

COMPASS-II will measure combined beam-spin and charge cross sections:
\begin{equation}
\begin{array}{lclcl}
\mathcal{S}_{\rm CS,U} 
& = & d\sigma^{\stackrel{+}{\rightarrow}} + d\sigma^{\stackrel{-}{\leftarrow}} 
& = & 2 d\sigma_{UU} ( 1 - A_{LU, I} ), \\
\mathcal{D}_{\rm CS,U}  
& = & d\sigma^{\stackrel{+}{\rightarrow}} - d\sigma^{\stackrel{-}{\leftarrow}} 
& = & 2 d\sigma_{UU} ( A_C - A_{LU, DVCS}  ), \\
\mathcal{A}_{\rm CS,U}  
& = &  \frac{d\sigma^{\stackrel{+}{\rightarrow}} - d\sigma^{\stackrel{-}{\leftarrow}}}{d\sigma^{\stackrel{+}{\rightarrow}} + d\sigma^{\stackrel{-}{\leftarrow}}} 
& = & \frac{A_C - A_{LU, DVCS}}{1 - A_{LU, I}}.
\end{array}
\end{equation}

These observables offer the interesting feature of  different sensitivity to various combinations of CFFs, as summarized in Tab.~\ref{tab:normalized-CFF-dependence}. In spite of the fact that $|t| / Q^2$ is not so small for  typical kinematics in Tab.~\ref{tab:Typical-kinematics-DVCS-measurements}, we do not take into account higher-twist contributions in our study, and restrict ourselves to the study of leading-order and next-to-leading order terms in the $\alpha_S$ expansion.

\begin{table}[t]
	\centering
	\begin{tabular}{ccc}
		\hline\noalign{\smallskip}
		\rule[-3mm]{0mm}{8mm}\bf Experiment   & \bf Observable	& \bf Normalized CFF dependence \\ 
		 [3pt] \tableheadseprule\noalign{\smallskip}
		 ~	& \rule[-2mm]{0mm}{7mm}$A_{\rm C}^{\cos 0\phi}$	& ${\rm Re} \mathcal H +0.06 {\rm Re} \mathcal E +0.24 {\rm Re} {\widetilde{\mathcal H}}$ \\ 
		HERMES	& \rule[-2mm]{0mm}{7mm}$A_{\rm C}^{\cos \phi}$	& ${\rm Re} \mathcal H +0.05 {\rm Re} \mathcal E +0.15 {\rm Re} {\widetilde{\mathcal H}} $ \\ 
         	~	& \rule[-2mm]{0mm}{7mm}$A_{\rm LU,I}^{\sin \phi}$	& ${\rm Im} \mathcal H +0.05 {\rm Im} \mathcal E +0.12 {\rm Im} {\widetilde{\mathcal H}} $ \\ 
         	~ & \rule[-2mm]{0mm}{7mm}$A_{\rm UL}^{+,\sin \phi}$	& ${\rm Im} {\widetilde{\mathcal H}} + 0.10{\rm Im} \mathcal H + 0.01{\rm Im} \mathcal E$ \\ 
		CLAS	& \rule[-2mm]{0mm}{7mm}$A_{\rm LU}^{-,\sin \phi}$	& ${\rm Im} {\mathcal H} + 0.06 {\rm Im} \mathcal E + 0.21 {\rm Im} {\widetilde{\mathcal H}}$ \\ 
		~	& \rule[-2mm]{0mm}{7mm}$A_{\rm UL}^{-,\sin \phi}$	& ${\rm Im} {\widetilde{\mathcal H}}  + 0.12{\rm Im} \mathcal H + 0.04{\rm Im} \mathcal E$ \\ 
		\noalign{\smallskip}\hline
	\end{tabular}
	\caption{Normalized dependence of various existing DVCS observables on the CFF at the kinematics specified in 
Tab.~\ref{tab:Typical-kinematics-DVCS-measurements}. The largest coefficient in front of a CFF is set to 1, and only relative coefficients larger than 1\% are kept. 
See Ref.~\cite{Kroll:2012sm} for more information.}
	\label{tab:normalized-CFF-dependence}
\end{table}

\begin{table}[t]
	\centering
	\begin{tabular}{ccccc}
		\hline\noalign{\smallskip}
		~  & \multicolumn{4}{c}{\bf Kinematics} \\ \cline{2-5} \raisebox{1.5ex}[0pt]{\bf Experiment}		  & \rule[-2mm]{0mm}{7mm}$x_B$	&	$Q^2$ [GeV$^2$]	&	  $t$ [GeV$^2$]	& $- t / Q^2$  \\
		 [3pt] \tableheadseprule\noalign{\smallskip}
		COMPASS		& 0.05	& 2.00	& -0.20 	& 0.10	\\ 
		HERMES		& 0.09	& 2.50	&-0.12 	& 0.05	\\ 
		CLAS		& 0.19	& 1.25	&-0.19 	& 0.15	\\ 
		\noalign{\smallskip}\hline
	\end{tabular}
	\caption{Typical kinematics of existing of near-future DVCS measurements.}
	\label{tab:Typical-kinematics-DVCS-measurements}
\end{table}

\begin{figure}
	\centering
	\includegraphics[width=0.5\textwidth]{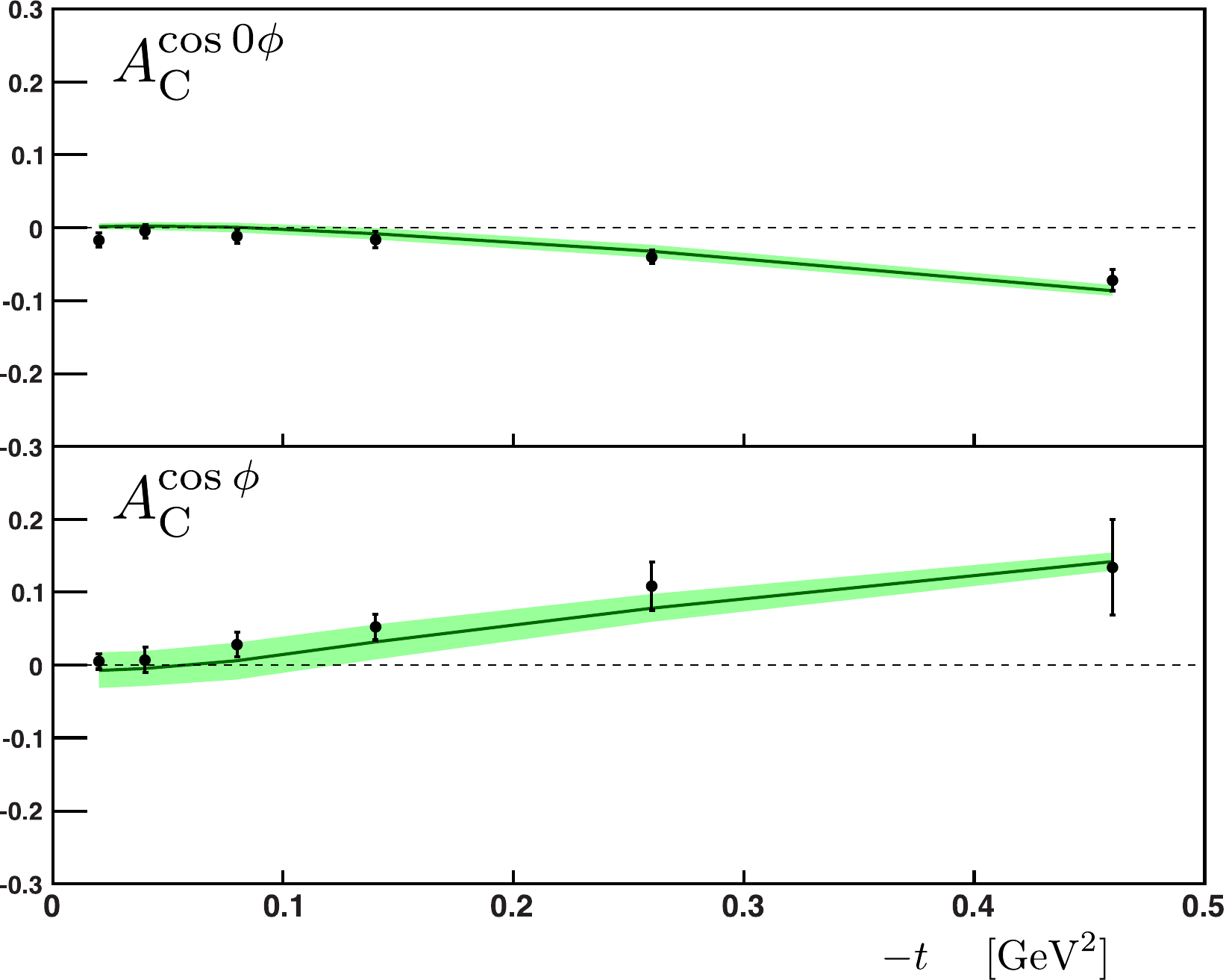}
	\caption{Beam Charge Asymmetry, HERMES \cite{Airapetian:2012mq}. See Ref.~\cite{Kroll:2012sm} for more information.}
	\label{fig:GK-HERMES-BCA}       
\end{figure}

\begin{figure*}
	\centering
	\includegraphics[width=0.75\textwidth]{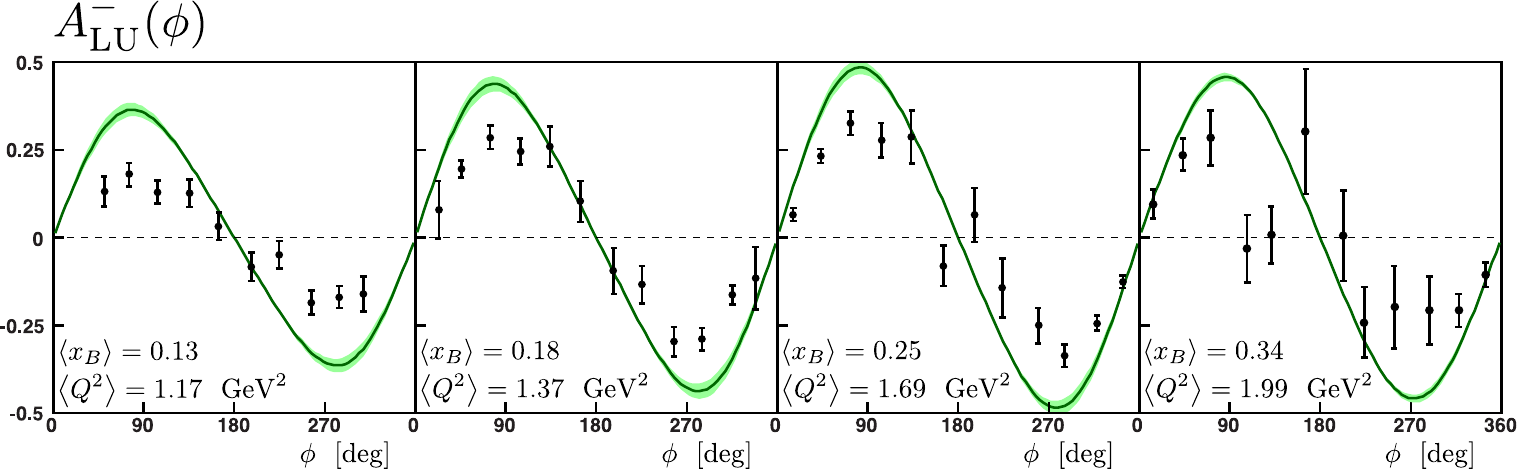}
	\caption{Beam Spin Asymmetry, CLAS \cite{Girod:2007aa}. See Ref.~\cite{Kroll:2012sm} for more information.}
	\label{fig:GK-CLAS-BSA}       
\end{figure*}

Figs.~\ref{fig:GK-HERMES-BCA} and \ref{fig:GK-CLAS-BSA} show the comparison of the predictions of the GK model (no parameter was tuned) to a selection of DVCS measurements. The model is in good agreement with the data, the agreement being better at small $\xi$ (which is the kinematic domain where the model was optimized for DVMP).

\subsection{Timelike and spacelike Compton Form Factors at Leading Order and Next to Leading Order}

We computed the real and imaginary parts of the CFF $\mathcal{H} = e_u^2 \mathcal{H}_u + e_d^2 \mathcal{H}_d + e_s^2 \mathcal{H}_s$ (where $e_q$ is the electric charge of the quark $q$ in units of $|e|$) at LO and NLO on a wide kinematic range (skewness varying between $10^{-3}$ and 1). Special attention was paid to the validation of the numerics: two independent codes were systematically compared and an accuracy of 0.1~\% has been achieved in the range under scrutiny.

\begin{figure*}
	\centering
	\includegraphics[width=0.75\textwidth]{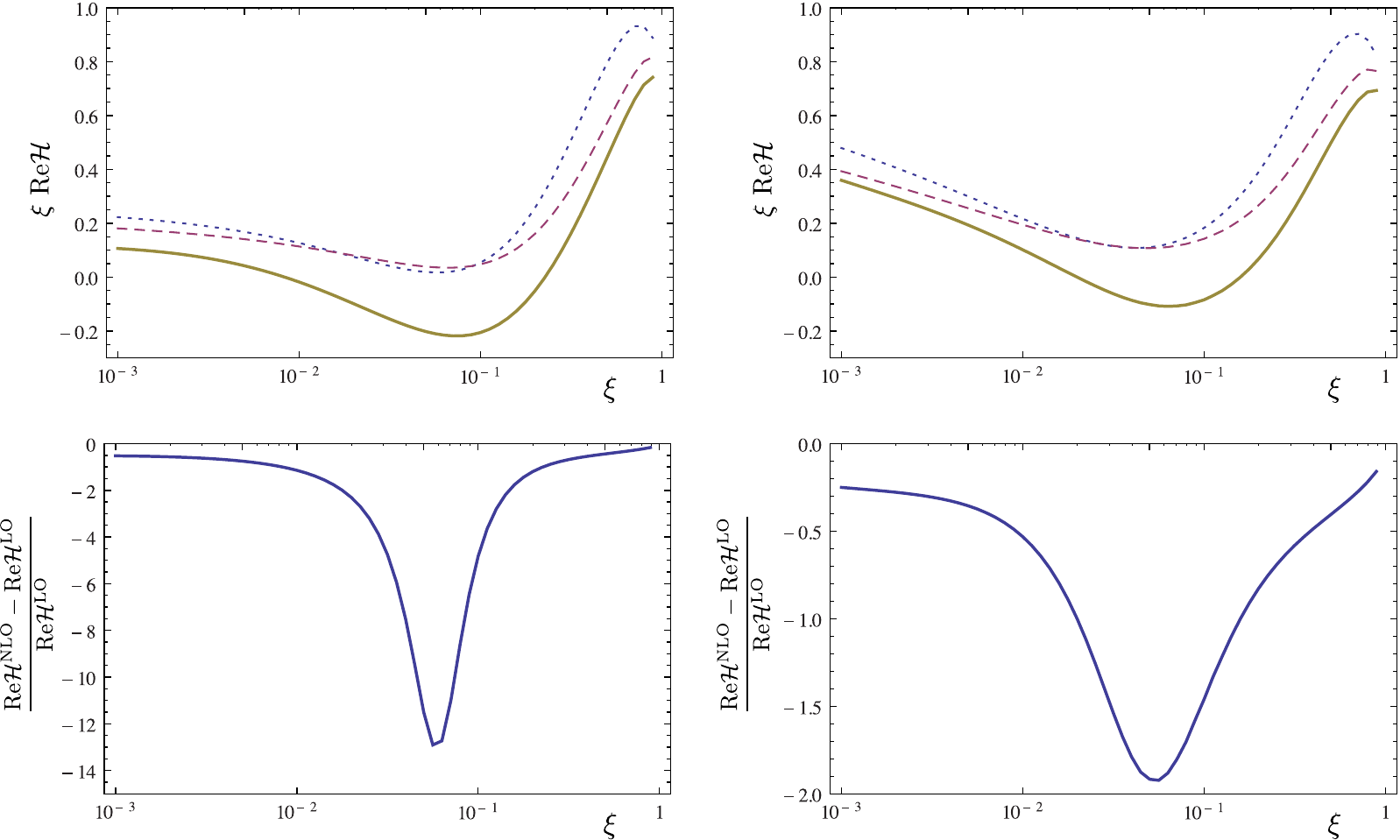}
	\caption{Spacelike $\ReH$ at LO and NLO ($t = -0.1~\GeV^2, Q^2 = \mu_F^2 = 4.~\GeV^2$). Blue dotted line: LO. Magenta dashed line: NLO quark corrections. Brown solid line: full NLO. Left: KG model. Right: MSTW08-based model. See Ref.~\cite{Moutarde:2013qs} for more information.}
	\label{fig:DVCS_Re}       
\end{figure*}

\begin{figure*}
	\centering
	\includegraphics[width=0.75\textwidth]{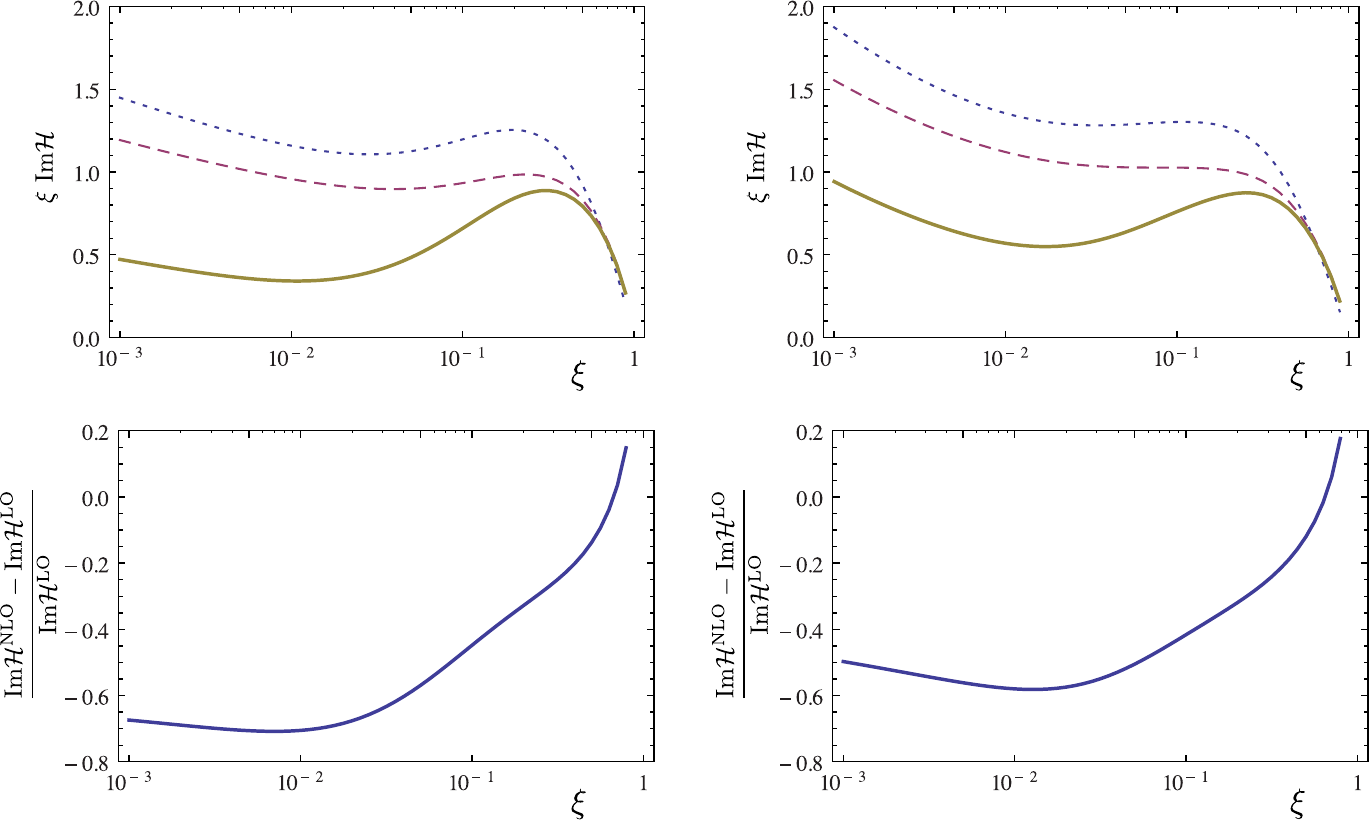}
	\caption{Spacelike $\ImH$ at LO and NLO ($t = -0.1~\GeV^2, Q^2 = \mu_F^2 = 4.~\GeV^2$). Blue dotted line: LO. Magenta dashed line: NLO quark corrections. Brown solid line: full NLO. Left: KG model. Right: MSTW08-based model. See Ref.~\cite{Moutarde:2013qs} for more information.}
	\label{fig:DVCS_Im}       
\end{figure*}

\begin{figure*}
	\centering
	\includegraphics[width=0.75\textwidth]{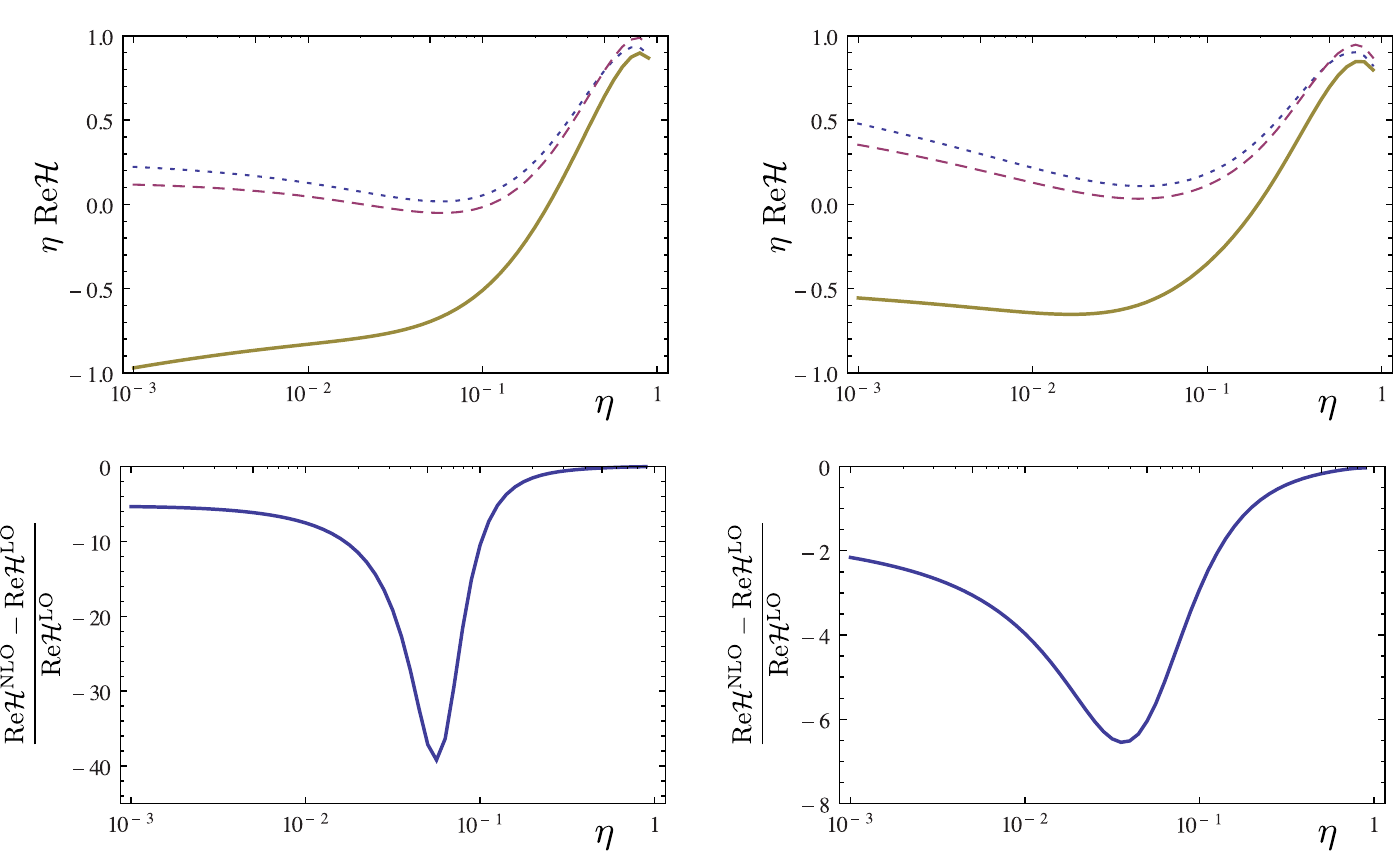}
	\caption{Timelike $\ReH$ at LO and NLO ($t = -0.1~\GeV^2, Q^2 = \mu_F^2 = 4.~\GeV^2$). Blue dotted line: LO. Magenta dashed line: NLO quark corrections. Brown solid line: full NLO. Left: KG model. Right: MSTW08-based model. See Ref.~\cite{Moutarde:2013qs} for more information.}
	\label{fig:TCS_Re}       
\end{figure*}

\begin{figure*}
	\centering
	\includegraphics[width=0.75\textwidth]{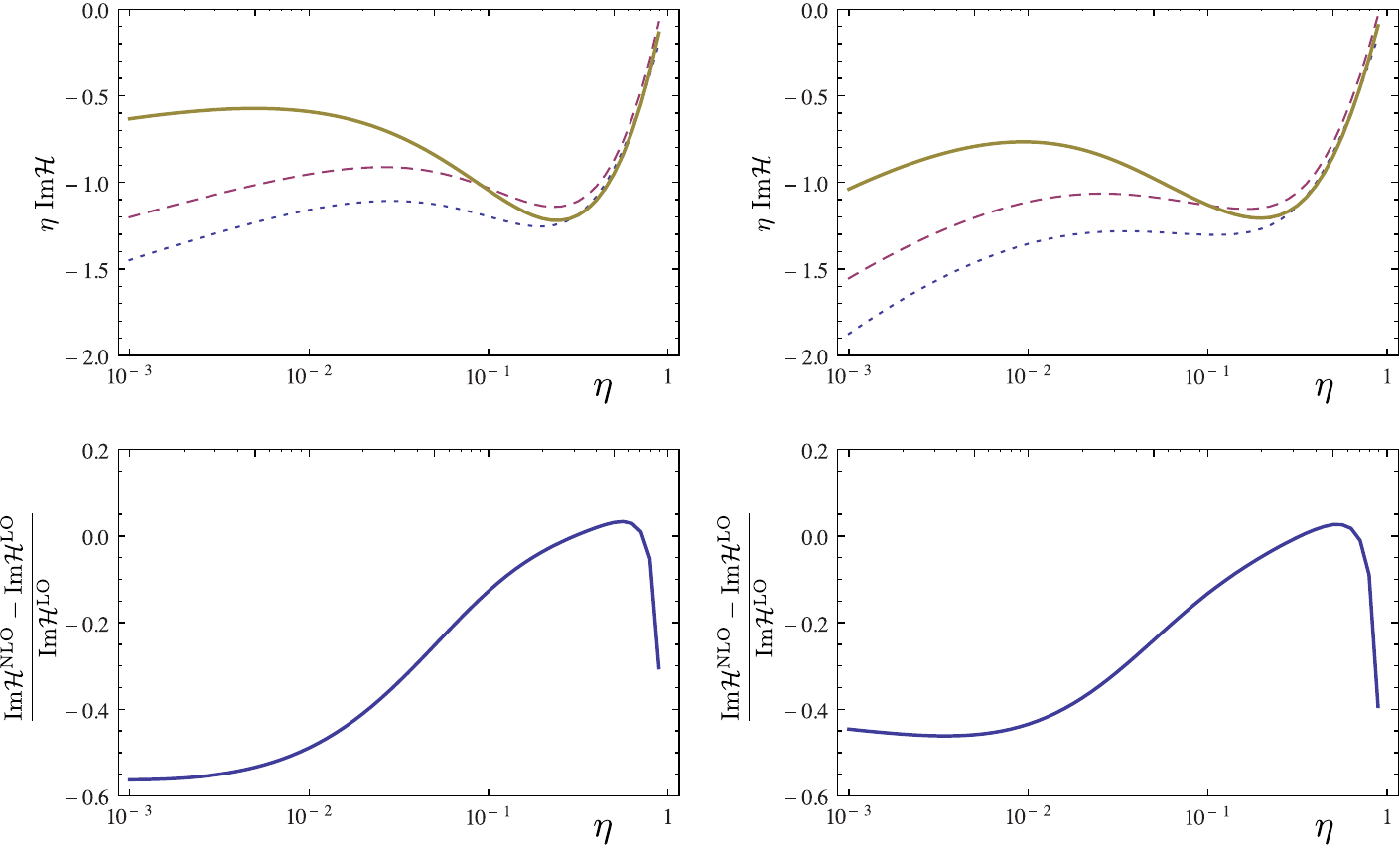}
	\caption{Timelike $\ImH$ at LO and NLO ($t = -0.1~\GeV^2, Q^2 = \mu_F^2 = 4.~\GeV^2$). Blue dotted line: LO. Magenta dashed line: NLO quark corrections. Brown solid line: full NLO. Left: KG model. Right: MSTW08-based model. See Ref.~\cite{Moutarde:2013qs} for more information.}
	\label{fig:TCS_Im}       
\end{figure*}

In all cases we have plot the LO CFF, and the NLO result with quark contributions only, or both quark and gluon contributions. The results are shown from Figs.~\ref{fig:DVCS_Re} to \ref{fig:TCS_Im}. We observe large NLO corrections, mostly due to gluon GPDs. Surprisingly these corrections are maximal in the kinematic region of HERMES and COMPASS. In all our calculations, we choose $\mu_F=Q$. The size of the corrections may depend on that choice, and this will be the subject of further studies. Also the choice of a particular scale needs some theoretical  justification.

\subsection{CLAS12}

\begin{figure*}
	\centering
	\includegraphics[width=0.75\textwidth]{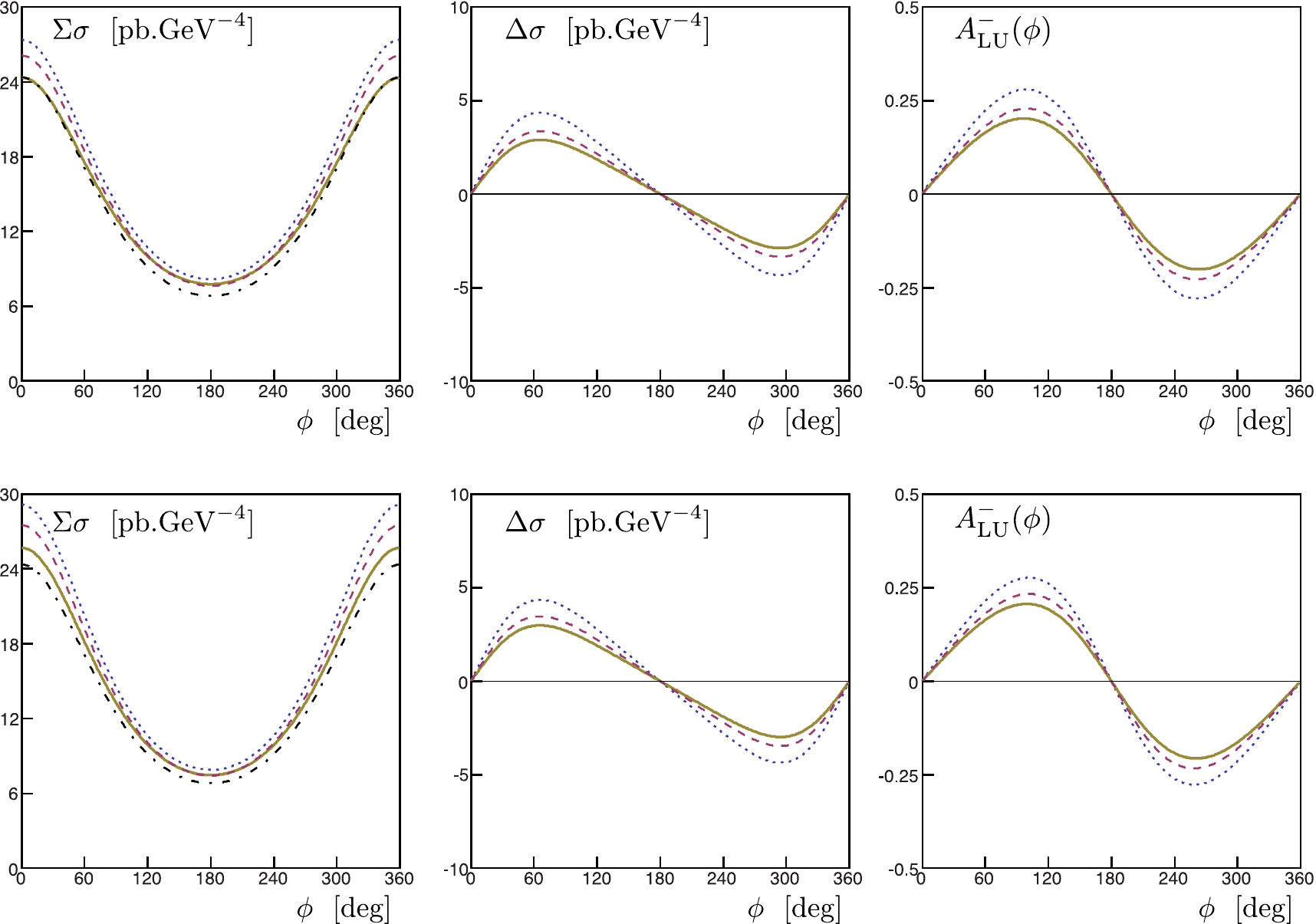}
	\caption{Prediction of beam-polarized and unpolarized DVCS cross sections for CLAS12 at LO and NLO ($E_e = 11.~\GeV, x_B = 0.36, t = -0.2~\GeV^2, Q^2 = \mu_F^2 = 4.~\GeV^2$) with the dominant CFF $\mathcal{H}$ only. Blue dotted line: LO. Magenta dashed line: NLO quark corrections. Brown solid line: full NLO. Upper line: KG model. Lower line: MSTW08-based model. See Ref.~\cite{Moutarde:2013qs} for more information.}
	\label{fig:CLAS12-DVCS-cross-section}       
\end{figure*}

The effects of NLO corrections shown in Fig.~\ref{fig:CLAS12-DVCS-cross-section} are quite large in both considered GPD models although the value of $\xi$ is rather large. In particular we see that the gluon contributions is by no means negligible.

\begin{figure*}
	\centering
	\includegraphics[width=0.5\textwidth]{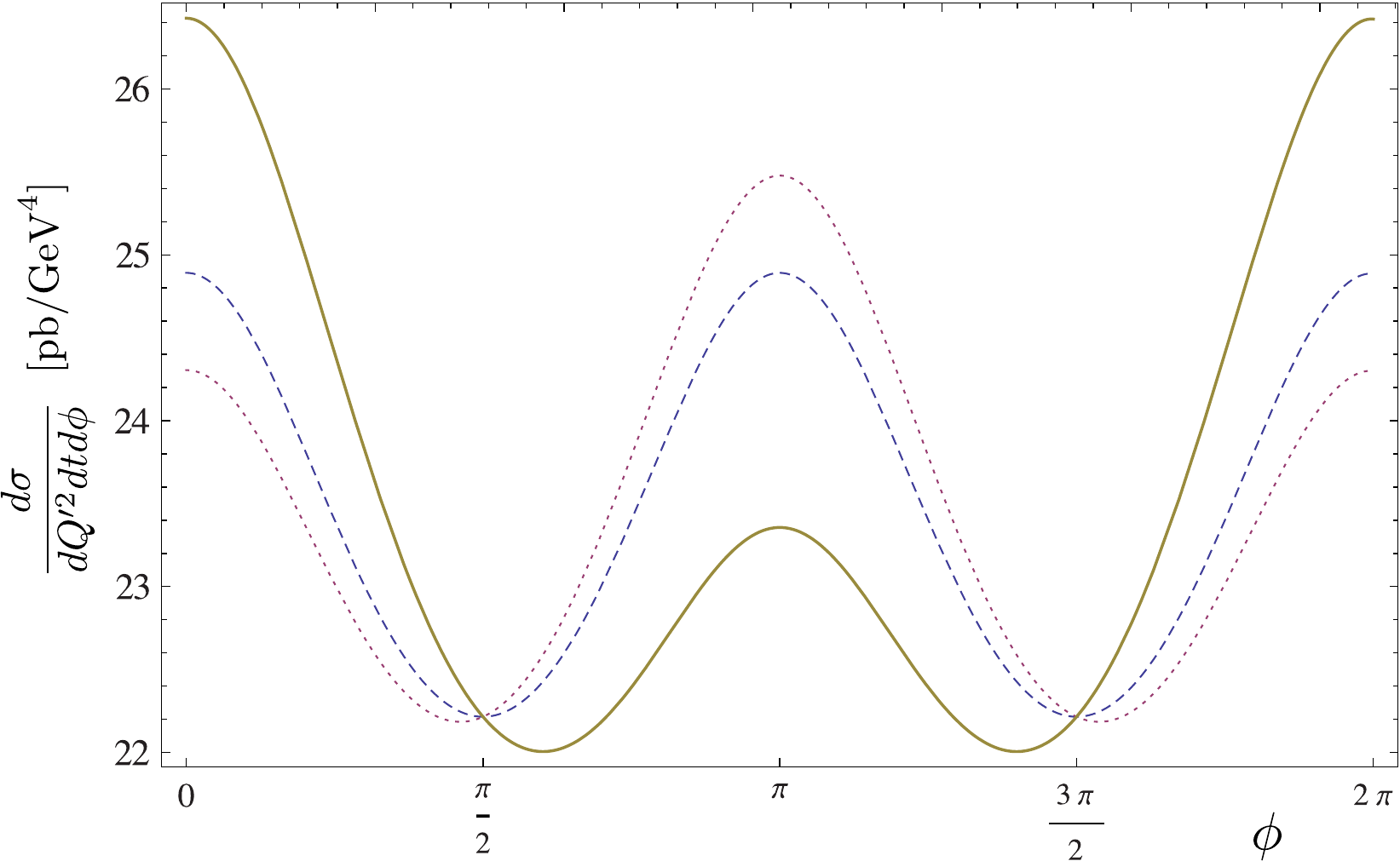}
	\caption{Prediction of TCS cross sections integrated over $\theta \in [ \pi / 4, 3 / 4 \pi ]$ for CLAS12 with the GK model at LO and NLO ($E_\gamma = 10.~\GeV (\eta \simeq 0.11), t = -0.1~\GeV^2, Q^2 = \mu_F^2 = 4.~\GeV^2$) with the dominant CFF $\mathcal{H}$ only. Blue dotted line: LO. Magenta dashed line: NLO quark corrections. Brown solid line: full NLO. See Ref.~\cite{Moutarde:2013qs} for more information.}
	\label{fig:CLAS12-TCS-cross-section}       
\end{figure*}

The $\phi$-dependence of the interference term allows for an access to the real part of the CFF $\mathcal{H}$. From Fig.~\ref{fig:TCS_Re} we know that this quantity is subject to large NLO corrections. On Fig.~\ref{fig:CLAS12-TCS-cross-section} we indeed compare the pure Bethe~-~Heitler term, and the pure Bethe~-~Heitler term augmented by its interference with TCS at LO and NLO. 

\subsection{COMPASS}

\begin{figure*}
	\centering
	\includegraphics[width=0.75\textwidth]{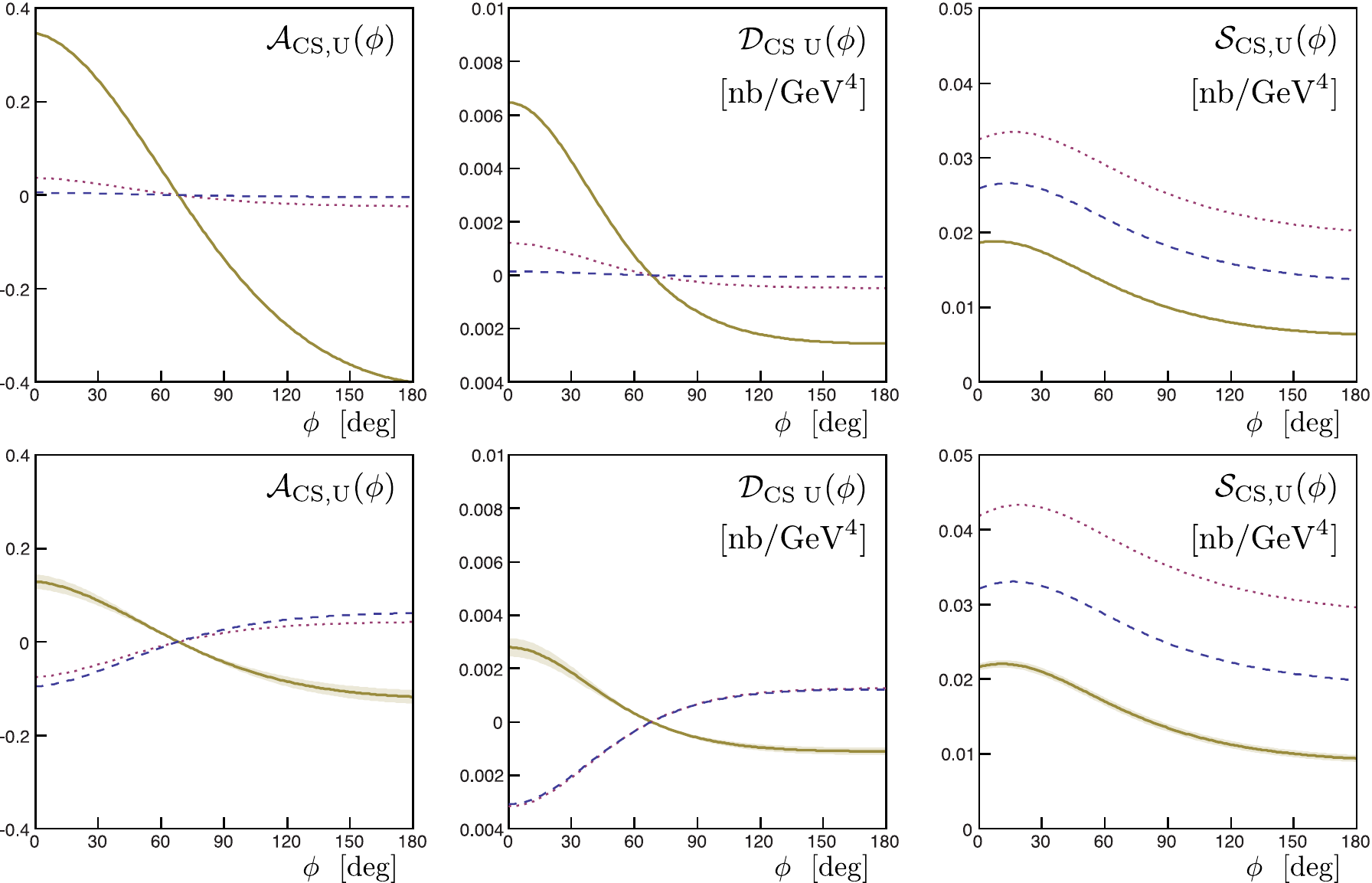}
	\caption{Prediction of combined charge-spin DVCS cross sections for COMPASS-II at LO and NLO ($E_\mu = 160.~\GeV, \xi = 0.05, t = -0.2~\GeV^2, Q^2 = \mu_F^2 = 4.~\GeV^2$) with the dominant CFF $\mathcal{H}$ only. Blue dotted line: LO. Magenta dashed line: NLO quark corrections. Brown solid line: full NLO. Upper line: KG model. Lower line: MSTW08-based model. See Ref.~\cite{Moutarde:2013qs} for more information.}
	\label{fig:COMPASS}       
\end{figure*}

From Fig.~\ref{fig:COMPASS} we observe that NLO effects should be highly visible at COMPASS which probes a kinematic region populated by sea quarks and gluons.

\section{Conclusions}

Deeply Virtual Compton Scattering, both in its spacelike and timelike realizations, is the golden channel to extract GPDs from measurements. Using model-dependent evaluations we have demonstrated here, in the case of medium energy kinematics which will be explored in the near future at JLab and COMPASS,  that the inclusion of NLO corrections to the coefficient function could be an important issue. The difference of these corrections between the spacelike and timelike regimes is so sizable that they can be promoted to the status of direct tests of the QCD understanding of the reactions. Considering CFF fitting, global fits of DVCS and TCS data will be needed to separate quark and gluon contributions and allow an accurate interpretation of extracted data. 

Finally,  let us emphasize that we do not consider as a weakness of the current physics program the apparent importance of NLO contributions to Compton form factors and DVCS or TCS observables. Although it certainly opens the way to a challenging  verification that NNLO corrections are either under control or subject to a legitimate resummation procedure \cite{Altinoluk:2012nt}, it points to the very positive fact that the COMPASS-II and JLab12 experiments may constrain gluon GPDs. The 3D tomography of the gluonic structure of the nucleon may thus be scrutinized through the $t$-dependence of Compton form factors extracted from near future experimental data.

\begin{acknowledgements}
We are grateful to M.~Diehl, N.~D'Hose, D.~M\"{u}ller, P.~Nadel-Turo\'nski,  W.-D.~Nowak, G.~Schnell, S.~Stepanyan and Samuel Wallon for many fruitful discussions and valuable inputs. 
\end{acknowledgements}



\end{document}